\documentclass[%
 preprint,
superscriptaddress,
 amsmath,amssymb,
]{revtex4-2}

\usepackage{ulem}
\usepackage[english]{babel}

\AtBeginDocument{\setcitestyle{numbers,round}} 

\usepackage{graphicx}
\usepackage{dcolumn}
\usepackage{bm}
\usepackage{color}
\DeclareUnicodeCharacter{2009}{\,}
\begin{document}
\setcitestyle{numbers,round} 
\renewcommand{\bibnumfmt}[1]{(#1)}
\renewcommand{\thetable}{\arabic{table}}
\begin{titlepage}
\centering

\vspace*{1cm}

{\Large\bfseries When and how particles are removed by drops\par}

\vspace{1.5em}

{\large
Abhinav Naga\textsuperscript{1,*,\ensuremath{\dagger}},
Franziska Sabath\textsuperscript{2,*},
Doris Vollmer\textsuperscript{2,\ensuremath{\dagger}},
Halim Kusumaatmaja\textsuperscript{1,\ensuremath{\dagger}}
\par}

\vspace{1em}

{\small 
\textsuperscript{1}Institute for Multiscale Thermofluids, School of Engineering, University of Edinburgh, Edinburgh, United Kingdom\\
\textsuperscript{2}Physics at Interfaces group, Max Planck Institute for Polymer Research, Mainz, Germany
}

\vspace{3em}

\begin{minipage}{\textwidth}
\small
\textsuperscript{*}These authors contributed equally.

\textsuperscript{\ensuremath{\dagger}}Corresponding authors. Emails: Abhinav Naga, \texttt{abhinav.naga@ed.ac.uk}; 
Doris Vollmer, \texttt{vollmerd@mpip-mainz.mpg.de}; 
Halim Kusumaatmaja, \texttt{halim.kusumaatmaja@ed.ac.uk}.
\end{minipage}

\maketitle

\end{titlepage}

\begin{center}
\textbf{Abstract}
\end{center}

\noindent
Particulate contaminants decrease the power output of solar panels, the transparency of windows, and are detrimental to microelectronics, where even a single particle can induce a short circuit. Despite significant research on particle adhesion and self-cleaning, it remains unclear when and how a drop can remove a particle from a surface, thus efficiently cleaning the surface. Here, by combining lattice Boltzmann simulations and confocal microscopy experiments, we show that at least six different scenarios arise from the complex interplay between capillary and friction forces when a drop collides with a particle on a surface. Notably, the capillary force plays a dual role in particle removal: while its tangential component always drives removal, its normal component can also hinder it. We introduce a dimensionless capillary capture parameter that successfully predicts particle removal across a wide range of particle and surface properties. Our results reveal that the design of easy-to-clean surfaces should not focus only on maximizing hydrophobicity, but also on minimizing the particle--surface friction.

\vfill

\section{Introduction}

\begin{figure*}
     \centering
     \includegraphics[width=0.78\linewidth]{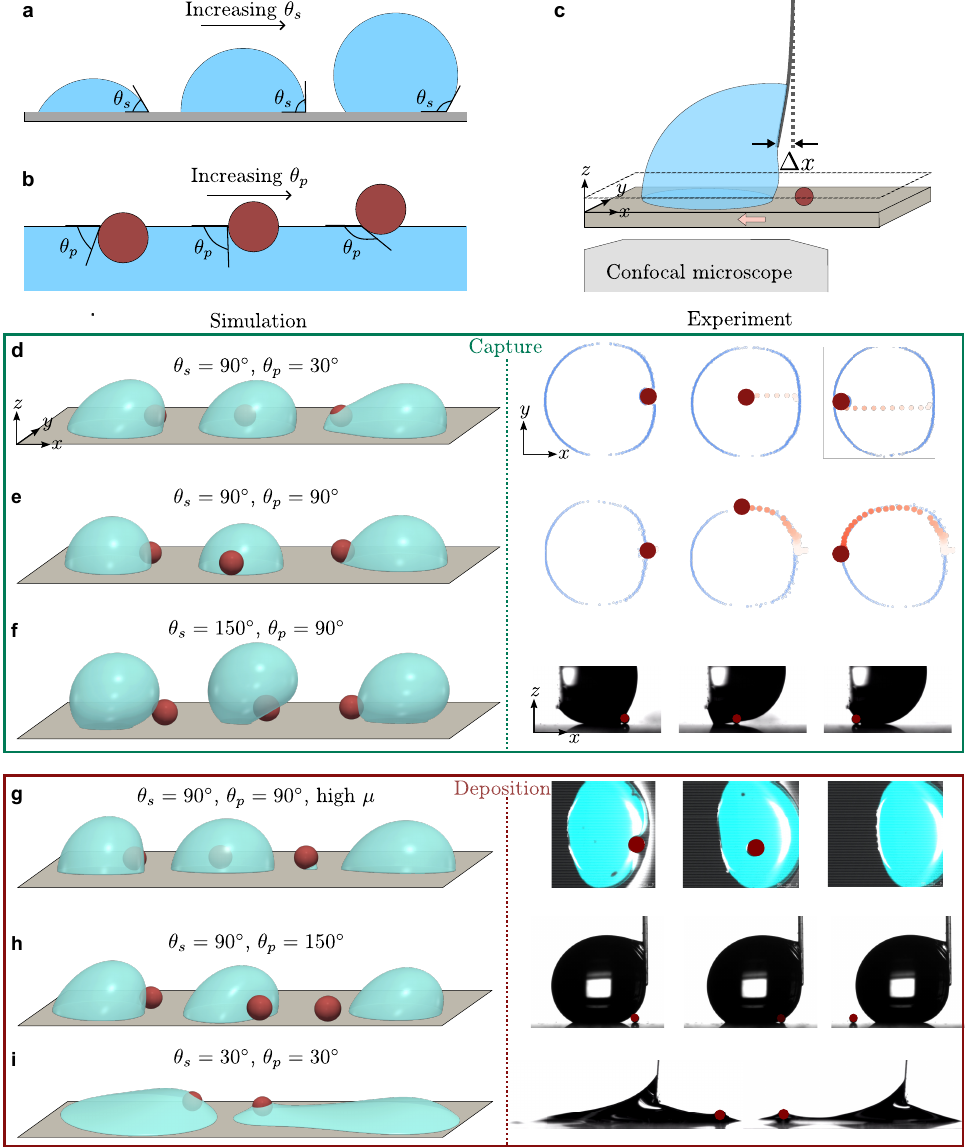}
     \caption{\textbf{Drop--particle interaction scenarios.} (a) Schematic of the equilibrium shape of a drop as a function of $\theta_s$. (b) Schematic of equilibrium position of a particle at a liquid-air interface as a function of $\theta_p$. (c) Experimental setup to image drop-particle interaction and to measure horizontal forces. {\color{black} (d--f) Interaction scenarios when the particle is captured by drop. (g--i) Interaction scenarios when the particle is redeposited by the drop. In (d--i), Left and right columns show simulations and experiments, respectively. In the simulations, the particle diameter was 200\,$\mu$m. Values of $\theta_p$ and $\theta_s$ listed for the simulations correspond to static contact angles. In the experiments, the particle diameter is in the range 210--250\,$\mu$m, except in (i) where it is between 355--420\,$\mu$m. In the experiments, $\theta_p$ ranges from $20^\circ$ to $145^\circ$ and $\theta_s$ ranges from $<30^\circ$ to $150^\circ$. See Table\,\ref{Table: Experimental details} for further experimental details. Right column of panel (g) is reproduced from Naga \textit{et al.} (2021)~\cite{naga_how_2021} under a Creative Commons CC BY 3.0 license.}}
     \label{fig:IntroFigure}
 \end{figure*}

{\color{black}The atmosphere contains a vast number of particles spanning several orders of magnitude in diameter, from a few nanometres to hundreds of micrometres~\cite{adebiyi_review_2023}, including mineral dust, microplastics, pollen, and other organic debris. These particles are continuously deposited onto natural and engineered surfaces. Once deposited, particles can severely affect the functionality of surfaces. On plant leaves, dust can impair photosynthetic activity~\cite{soheili_effect_2023,beckett_urban_1998}. On photovoltaic panels, dust reduces light absorption and therefore power output, leading to annual economic losses of more than 3 billion euros globally~\cite{ilse_techno-economic_2019}. In high-precision industries, such as semiconductor manufacturing, particulate contamination is so costly that extensive resources are devoted to preventing deposition in the first place by operating in expensive cleanrooms~\cite{plummer_semiconductor_2023}. However, on outdoor surfaces, the accumulation of contaminant particles is unavoidable and we must therefore develop strategies to clean them sustainably. 

Existing strategies for cleaning surfaces include applying liquids containing chemical detergents, mechanical wiping, applying adhesive peel-off tapes, vacuuming, or repelling particles using electric fields~\cite{qin_mechanisms_2003, parrott_automated_2018, sayyah_mitigation_2013, wilson_effectiveness_2017, panat_electrostatic_2022, yu_polymer-based_2023}. However, these approaches can be chemically harmful, mechanically damaging, energy-intensive, costly, or impractical over large areas. In contrast, a promising strategy is to develop surfaces that are inherently easy to clean, so that particles are removed naturally by raindrops sliding across the surface. Prominent types of surfaces with such easy-to-clean properties include superhydrophobic surfaces~\cite{neinhuis_characterization_1997,blossey_self-cleaning_2003, hassan_self-cleaning_2019, furstner_wetting_2005, lafuma_superhydrophobic_2003,wang_self-cleaning_2025,bhushan_self-cleaning_2009}, liquid-infused surfaces~\cite{wong_bioinspired_2011,david_smith_droplet_2013, quere_wetting_2008, naga_direct_2024} and liquid-like surfaces~\cite{fadeev_trialkylsilane_1999, krumpfer_contact_2010, chen_omniphobic_2023, gresham_comparing_2025}. These surfaces differ in their wettability, with water drops adopting a wide range of contact angles ($\theta_s$, Fig.\,\ref{fig:IntroFigure}a) ranging from around $90^\circ$ to values approaching $180^\circ$~\cite{naga_drop_2025,hauer_wetting_2024, daniel_origins_2018}. At the same time, particulate contaminants present in the environment also span a broad range of wettabilities, from relatively hydrophilic mineral dust~\cite{deng_direct_2018} to hydrophobic microplastics~\cite{al_harraq_microplastics_2021, prajapati_microplastic_2022}. Therefore, there are many possible combinations of surface and particle wettability (Fig.\,\ref{fig:IntroFigure}b), yet it remains unclear how these different combinations influence whether a particle can be removed by a drop sliding across the surface. In particular, the complex interplay between capillary, hydrodynamic, frictional, and adhesion forces has so far precluded the development of a quantitative framework for particle removal.

Drops offer an effective way to remove particles due to capillarity. When a drop encounters a particle on a surface, the liquid--air interface deforms and exerts a capillary force $F_\gamma \sim \gamma R_p$ on the particle, where $\gamma$ is the liquid surface tension and $R_p$ is the particle radius~\cite{naga_how_2021,geyer_when_2020,heckenthaler_self-cleaning_2019}. The particle also experiences a hydrodynamic viscous force due to shear in the liquid. The viscous force scales as $F_\eta \sim \eta_d (V/H)R_p^2$, where $\eta_d$ is the dynamic drop viscosity, $V$ is the drop velocity, and $H$ is the drop height~\cite{butt_surface_2018}. For submillimetric particles, the capillary force generally dominates, $F_\gamma/F_\eta \sim \gamma H/(\eta_d V R_p)\gg1$, even for drop velocities as large as several cm/s. For even smaller particles, the capillary force becomes increasingly more dominant because it decreases only linearly with $R_p$, whereas the viscous force decreases more rapidly with $R_p^2$. Capillary forces therefore offer a powerful route for removing small particles. Using capillary forces to remove particles also has the advantage that it is effective with pure water, thus reducing the need for surfactants that are typically used during cleaning.

For a particle to be removed, the capillary force must overcome the forces holding the particle on the substrate, particularly adhesion~\cite{heckenthaler_self-cleaning_2019,leenaars_particle_1989, geyer_when_2020, perumanath_contaminant_2023,butt_surface_2018} and friction~\cite{naga_how_2021,yin_dynamic_2018}. Classical analytical models~\cite{scheludko_measurement_1975} that predict the capillary force between a particle and a liquid--fluid interface are not designed to account for the presence of a solid surface. This omission is clearly a major limitation in the context of particle removal because even from our everyday experience, it is clear that some surfaces are easier to clean than others. Therefore, models for predicting particle removal must not treat the process merely as two-body particle--drop problem, but rather as a three-body particle--drop--surface problem.

Due to the lack of sufficient spatial and/or temporal resolution needed to study particle removal in detail, existing experimental studies are either qualitative or have mainly focused on quantifying the total number of particles on the surface before and after the passage of a drop without resolving the interaction process~\cite{neinhuis_characterization_1997,blossey_self-cleaning_2003, hassan_self-cleaning_2019, furstner_wetting_2005, lafuma_superhydrophobic_2003,wang_self-cleaning_2025, bhushan_self-cleaning_2009}. Furthermore, experimentally, only the horizontal force component in the direction of motion can be measured, which is insufficient to predict particle removal. Numerical studies of particle removal would be highly valuable to understand the underlying mechanism, but existing computational methods cannot usually account for all relevant forces, in particular capillary and friction forces. Recent developments in coupling the lattice Boltzmann method and the discrete element method~\cite{naga_modeling_2025, fei_coupled_2023, jiang_coupled_2022} have opened up the possibility to develop a quantitative framework for understanding particle removal.

Here, we reveal that particle removal is governed by a coupled capillary--frictional mechanism. To investigate this problem, we combine simulations using a state-of-the-art lattice Boltzmann--discrete element computational method~\cite{naga_modeling_2025} with experiments using confocal microscopy. Together, these methods provide the time-dependent drop shape, particle trajectory, and forces during the interaction. We observed six different collision pathways. A particle that is eventually captured by the drop can travel (1) inside, (2) underneath, or (3) to the side of the drop before ending up at the rear of the drop. A particle that is eventually left behind can also travel (4) inside or (5) to the side of the drop but then detach from the drop, or it can (6) remain in a water film behind the drop.

We show that the magnitude of the capillary force is primarily determined by the particle wettability, but its orientation depends on both the particle wettability and the surface wettability. In particular, the vertical component of the capillary force has a dual role: it can lift the particle and reduce friction, promoting removal, or press the particle against the substrate and increase friction, preventing removal. To predict whether a particle can be captured, we introduce a dimensionless capillary capture parameter that successfully describes particle removal across a broad range of particle and surface properties. Our results reveal previously underappreciated strategies for designing easy-to-clean surfaces. The surface needs to be sufficiently hydrophobic so that the receding side of the drop maintains its shape rather than elongating into a residual liquid film. However, rather than only maximizing surface hydrophobicity, it is crucial to also minimize particle--surface friction.}

\section{Results and discussion}

\subsection{Overview of collision scenarios}

{\color{black} The wettabilities of the particle and the surface play a crucial role in the mechanism of particle removal. To highlight the importance of wettability, we investigated the collision between a water drop and different types of particles on a variety of surfaces. Our experiments and simulations cover a wide range of surface contact angles $\theta_s$ (Fig.\,\ref{fig:IntroFigure}a) and particle contact angles $\theta_p$ (Fig.\,\ref{fig:IntroFigure}b), ranging from $\approx20^\circ-150^\circ$.

Experimentally, we directly imaged the collision from below using an inverted confocal microscope equipped with a cantilever (Fig.\,\ref{fig:IntroFigure}c, as introduced in~\cite{naga_how_2021}). We kept the drop in place using a stainless steel cantilever while moving the surface at controlled speeds (100\,$\mu$m/s--1\,mm/s). This technique provides bottom view image of the drop shape and the particle trajectory throughout the collision without the drop moving out of the limited field of view ($\approx 1$\,mm) of the microscope. Furthermore, this setup also allows us to measure the total horizontal force acting on the drop and the particle along $x$. The total force can be obtained using Hooke's law, $F_\mathrm{tot}=k\Delta x$, where $k$ is the spring constant of the cantilever and $\Delta x$ is the deflection (Fig.\,\ref{fig:IntroFigure}c). We verified that the local drop deformation induced by the cantilever does not influence our results by showing that the drop--particle collision scenario is independent of whether the cantilever is positioned in front or at the rear of the drop (Supplementary Section II and Supplementary Figure S1). To complement the bottom view imaging with a side view perspective, we also performed experiments with a goniometer to image the collision from the side. We conducted experiments with water drops with volumes ranging from 2--10\,$\mu$L and particle diameters ranging from 210--420\,$\mu$m. In all cases, the drop length was several times larger than the particle diameter.

We identified six collision scenarios when conducting experiments with a variety of particles and surfaces. These scenarios can be summarized as follows: 
\begin{enumerate}

    \item The particle entered the drop, traveled across the drop base, and remained attached at the rear (Fig.\,\ref{fig:IntroFigure}d).
    
    \item The particle stayed at the drop--air interface, moved around the drop perimeter, and remained attached at the rear (Fig.\,\ref{fig:IntroFigure}e).

    \item The particle stayed at the drop--air interface, but unlike the above scenario, it traveled under the drop rather than on the side to reach the rear (Fig.\,\ref{fig:IntroFigure}f).

    \item The particle entered the drop, traveled across the drop base and remained attached at the rear side (Fig.\,\ref{fig:IntroFigure}g).
    
    \item The particle moved around the front half of the drop and eventually detached when it reached the widest part of the drop (Fig.\,\ref{fig:IntroFigure}h).
    
    \item The particle entered or moved around the drop and was then deposited in a water film behind the drop (Fig.\,\ref{fig:IntroFigure}i).
    
\end{enumerate}
These six scenarios can be divided into two main categories: capture (first three scenarios, Fig.\,\ref{fig:IntroFigure}d--f) and deposition (last three scenarios, Fig.\,\ref{fig:IntroFigure}g--i). Capture refers to when the particle remains attached to the drop and is carried away, whereas deposition refers to the case where the particle is left behind.

To gain insight into when each scenario occurs, we simulated the drop--particle collision numerically using our LBM-DEM framework. These simulations allow us to investigate the effects of each key parameter, including $\theta_p$, $\theta_s$, and the particle--surface friction coefficient $\mu$, independently. Typically, it is not possible to decouple the effects of these parameters in experiments. For example, varying $\theta_s$ requires changing the surface chemistry or topography, both of which also cause the particle adhesion and friction coefficient to change.  Our numerical method allows us to control all these parameters independently, providing insight into how each parameter affects the collision scenario.

In the simulations, we moved the drop by directly applying a force to it. The simulations were conducted in the same regime as the experiments, where capillary forces between the drop and the particle exceed hydrodynamic viscous forces (see Materials and Methods and Supplementary Table S2 for details of simulation parameters). With the simulations, we can visualize the three-dimensional drop shape (left column of Fig.\,\ref{fig:IntroFigure}d--i, Supplementary Movies 2--7) and obtain both the horizontal and vertical forces acting on the particle as it interacts with the drop. As we will discuss later in this paper, the vertical force component---which cannot be obtained in the experiments---plays an important role in determining whether a particle can be captured. In the following, we introduce a framework to rationalize the different scenarios and to predict the final outcome in terms of the forces acting on the particle.}

\subsection{Framework for particle capture}

\begin{figure}[h!]
     \centering
     \includegraphics{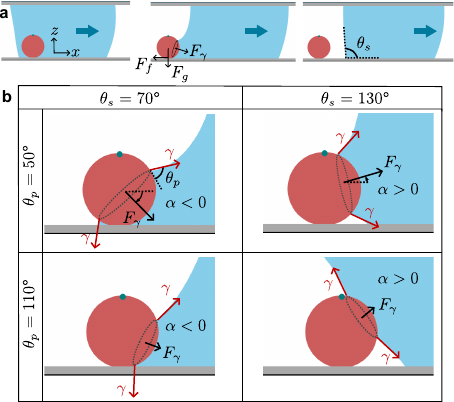}
     \caption{\textbf{Forces acting on particle.} (a) Simulation snapshots of liquid-air interface moving across a particle. The contact angle on the top surface was set to ($180^\circ-\theta_s$) so that the interface is initially straight to replicate the scenario where a small particle interacts with a large drop. (b) Magnified simulation snapshots showing the shape of the capillary bridge and the orientation of the capillary force for different $\theta_p$ and $\theta_s$. The total capillary force $F_\gamma$ was obtained by integrating the surface tension vector (red arrows) along the three-phase contact line (dashed circle in gray). The magnitude of $F_\gamma$ depends mainly on $\theta_p$ whereas its orientation $\alpha$ depends on both $\theta_s$ and $\theta_p$. $\alpha$ is the angle between the capillary force and the horizontal.}
     \label{fig:CapillaryBridge}
 \end{figure}

To understand the mechanism governing particle capture, we consider the forces acting on the particle when it is attached to the rear drop-air interface. We focus on the rear interface because when the particle is captured (Fig.\,\ref{fig:IntroFigure}d,e,f), it always ends up at the rear side. For liquids having a low viscosity (\textit{e.g.,} water with dynamic viscosity $\eta_d=1\,\mathrm{mPa\,s}$), the capillary force ($F_\gamma\sim \gamma R_p$) exerted by the drop on the particle dominates the viscous force ($F_\eta\sim \eta_d (V/H) R_p^2$, where $V$ is the drop speed and $H$ is the drop height) for speeds up to a few m/s~\cite{naga_how_2021}. We expect the particle to remain attached to the receding drop-air interface when
\begin{equation}
F_\gamma^x >F_f,
\label{Eq. Criterion}
\end{equation}
where $F_\gamma^x$ is the component of the capillary force along the direction of motion $x$, and $F_f$ is the \textcolor{black}{total resistive} force that opposes the motion of the particle. {\color{black}In general, the line of action of the capillary force and the friction force do not intersect (middle panel of Fig.\,\ref{fig:CapillaryBridge}a). Therefore, these forces also generate a torque on the particle. For example, in Fig.\,\ref{fig:CapillaryBridge}a (middle panel), friction at the contact point generates a clockwise torque about the particle's center of mass.}

{\color{black}In general, the resistive force $F_f$ that should be included in Eq.\,\ref{Eq. Criterion} depends on whether the particle slides or rolls. When the particle slides, $F_f$ corresponds to the sliding friction. In contrast, when the particle rolls, $F_f$ should include the effective force contributions due to rolling friction and resistive capillary torque~\cite{naga_how_2021, naga_capillary_2021, marshall_capillary_2014}. In the following, we introduce a framework for predicting when a particle is captured, focusing on sliding friction because sliding is the predominant mode of motion in our current simulations. In the Supplementary Information, we discuss how to extend the framework to account for rolling. Our LBM-DEM method can account for both sliding and rolling motion and it will be interesting to explore the relative importance of sliding versus rolling in the future. } 

The {\color{black}sliding} friction force is related to the normal force on the particle according to Amonton's law~\cite{gao_frictional_2004,butt_surface_2018},
\begin{equation}
F_f=\mu(F_p-F^z_\gamma),
\label{Eq. Amonton law}
\end{equation}
where $\mu$ is the coefficient of sliding friction, $F_p$ is the sum of the vertical forces (\textit{e.g.,} gravitational force and particle--surface adhesion) on the particle and $F_\gamma^z$ is the component of the capillary force {\color{black}due to the drop} in the vertical (upward) direction.

In general, the contact force $F_p$ can include different contributing factors. For small neutrally charged particles (radius $R_p\ll1\,\mu$m), van der Waals forces dominate the interaction. For large neutrally charged particles ($R_p\gg 100\,\mu$m), the gravitational force becomes more dominant than van der Waals forces because the gravitational force scales with $R_p^3$ whereas van der Waals forces typically scale only with $R_p$~\cite{butt_surface_2018}. For particles in between these extremes, both forces contribute. When the particle and the surface are charged, the electrostatic force becomes significant and can dominate even for large particles. Unlike van der Waals and the gravitational forces, the electrostatic force can either increase (attraction) or reduce (repulsion) the contact force. {\color{black}It is crucial to quantify the relevant interactions that influence $F_p$ and $F_f$ in the liquid (drop) because the interactions can vary significantly in air and water. In particular, in water, particles and surfaces that are neutral in air can become charged and develop an electric double layer that would not be present in air~\cite{butt_surface_2018,gu_potential_2000,liu_detecting_2020}. The presence of electric double layers can significantly influence both $F_p$ and $\mu$~\cite{israelachvili_intermolecular_2011}. Furthermore, the contribution due to van der Waals forces between two solids is also different when the two solids interact in water rather than air~\cite{butt_surface_2018}.}

In this work, since the particle is always in contact with the surface, the origin of $F_p$ is not important. What is important is the relative size of the friction force and the capillary force. Thus, for simplicity, we only included the gravitational force ($\approx4/3\pi R_p^3(\rho_p-\rho_l)g$, where $(\rho_p-\rho_l)$ is the density difference between the particle and the liquid, and $g$ is the gravitational acceleration) in our numerical model because the gravitational force is the dominant contribution for the large particles ($R_p> 100\,\mu$m) used in the experiments.  This simplification does not compromise the generality of our findings. The framework we propose in this paper remains valid even if $F_p$ is dominated by van der Waals or electrostatic forces because the friction force in Eq.\,\ref{Eq. Amonton law} is agnostic to whether $F_p$ originates from gravitational, van der Waals, or electrostatic forces.

By combining Eqs.\,\ref{Eq. Criterion} and \ref{Eq. Amonton law}, we obtain the following criterion for a successful capture:
\begin{equation}
F_\gamma^x >\mu(F_p-F^z_\gamma).
\label{Eq. Capture inequality}
\end{equation}
Equation\,\ref{Eq. Capture inequality} highlights that the horizontal force component $F_\gamma^x$ plays a direct role in overcoming the friction force. In contrast, the vertical component $F_\gamma^z$ can promote or hinder particle capture depending on whether it is positive (points upward) or negative (points downward).

\subsection{Capillary force on particle}

To understand how the magnitude and orientation of the capillary force depend on the particle wettability and surface wettability, we simulated a capillary bridge moving across a fixed spherical particle (Fig.\,\ref{fig:CapillaryBridge}a, see Materials and Methods for justification of simulation setup). When the particle emerged from the liquid, a three-phase contact line formed around the particle. Surface tension acts at the contact line, exerting a capillary force
\begin{equation}
    \vec{F_\gamma}=\oint_\mathrm{CL}\vec{\gamma}.\mathrm{d}\vec{l},
    \label{Eq. Capillary force integral}
\end{equation}
 where $\vec{\gamma}$ is the surface tension vector and the line integral is performed around the contact line. Although Eq.\,\ref{Eq. Capillary force integral} is an exact equation for the capillary force, an analytical expression for $\vec{F_\gamma}$ in terms of the contact angles $\theta_p$ and $\theta_s$ is not available. Our simulations allow us to evaluate the integral numerically for a range of $\theta_p$ and $\theta_s$. These simulations highlight that both $\theta_s$ and $\theta_p$ influence the capillary force.
 
 Increasing $\theta_s$ rotates the capillary force vector upward, causing $F_\gamma^z$ to become less negative and even switch sign at high contact angles, as shown in Fig.\,\ref{fig:CapillaryBridge}b. Similarly, increasing $\theta_p$ also rotates the capillary force vector upward. However, increasing $\theta_p$ also decreases the maximum value reached by the capillary force as the contact line moves across the particle, as indicated by the length of the $F_\gamma$ arrow in Fig.\,\ref{fig:CapillaryBridge}b. The maximum capillary force is lower for higher $\theta_p$ because the circumference of the contact line decreases with increasing $\theta_p$, causing the surface tension vector to act over a shorter length.

\begin{figure*}[h!]
     \centering
     \includegraphics{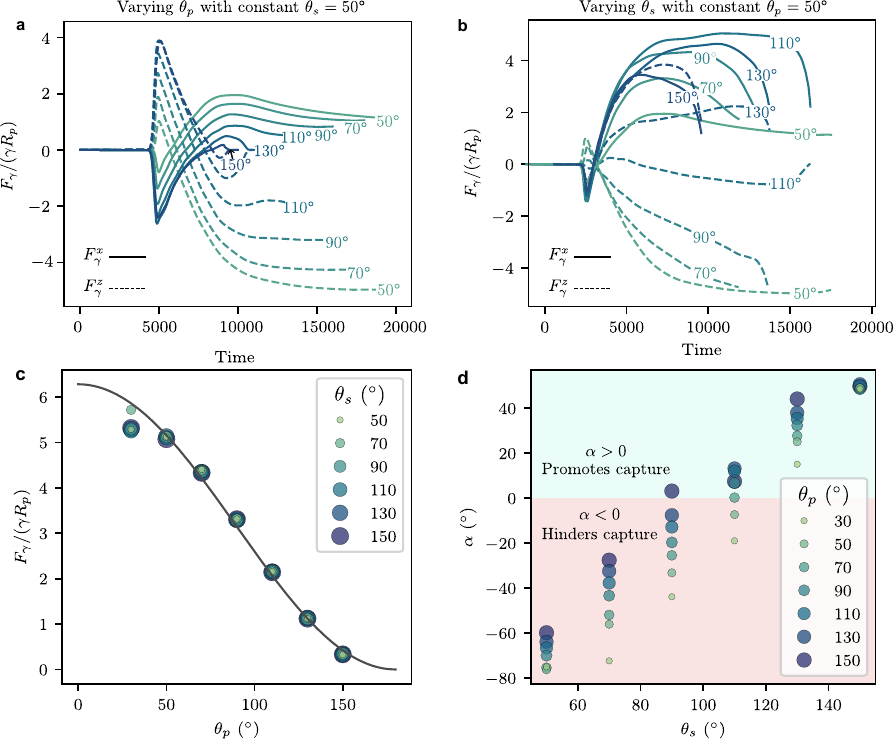}
     \caption{\textbf{Capillary force on particle as a function of particle wettability and surface wettability.} (a,b) Normalized capillary force on a particle as a water-air interface moved across the particle (as shown in Fig.\,\ref{fig:CapillaryBridge}a) as a function of the simulation time step. Solid and dashed lines correspond to the horizontal and vertical components of the capillary force, respectively. The line color represents different contact angles (green to blue represents increasing angle). In (a), $\theta_s$ is kept constant at $50^\circ$ while varying $\theta_p$. In (b) $\theta_p$ is kept constant at $50^\circ$ while varying $\theta_s$. Force curves for other values of $\theta_p$ and $\theta_s$ are provided in Supplementary Fig.\,S2. (c) Magnitude of the maximum capillary force as a function of $\theta_p$. The magnitude of the maximum capillary force depends predominantly on $\theta_p$ and barely varies with $\theta_s$. The equation $F_\gamma^\mathrm{max}= 2\pi R_p \gamma \cos^2(\theta_p/2)$ provides a good fit to the data, as shown by the black line. (d) Orientation of the capillary force (defined by the angle $\alpha$, Fig.\,\ref{fig:CapillaryBridge}a) for different $\theta_p$ and $\theta_s$ at the instant when the magnitude of the capillary force is maximum. $\alpha$ increases with both $\theta_s$ and $\theta_p$. }
     \label{fig:ForceCurves}
 \end{figure*}

To investigate how the horizontal and vertical components of the capillary force evolve, we plotted both components ($F_\gamma^x$ and $F_\gamma^z$, respectively) over time as a liquid-air interface moved across the particle (Fig.\,\ref{fig:ForceCurves}a,b). First, we investigate the effect of varying $\theta_p$ at a fixed $\theta_s=50^\circ$ (Fig.\,\ref{fig:ForceCurves}a). Initially, the particle was fully immersed in the liquid (Fig.\,\ref{fig:CapillaryBridge}a) and therefore the capillary force starts at zero. When the liquid-air interface came into contact with the particle, the force curves showed a sharp snap-in force. This snap-in corresponds to liquid dewetting the surface of the particle to establish the prescribed contact angle. For $\theta_s<90^\circ$, the snap-in force for $F_\gamma^z$ (dashed lines) can be strongly positive, which can help overcome adhesion forces between the particle and the substrate.

The shape of the force curves beyond the point where the snap-in force relaxes back to zero corresponds to the capillary force acting on the particle as it detaches from the interface. During detachment, $F_\gamma^x$ is always positive (solid lines), pulling the particle to the right. At constant $\theta_s$, the force curves shift toward zero with increasing $\theta_p$. Both $F_\gamma^x$ and $F_\gamma^z$ decrease with increasing $\theta_p$ due to the decrease in the length of the three-phase contact line.

Increasing $\theta_s$ while keeping $\theta_p$ constant shifts the curves corresponding to $F_\gamma^z$ upward, causing the force to switch from negative to positive above $\theta_s=110^\circ$ (Fig.\,\ref{fig:ForceCurves}b). The change in sign is due to the overall shape of the liquid-air interface switching from a downward-facing to an upward-facing configuration, as shown in Fig.\,\ref{fig:CapillaryBridge}b (top row). In contrast, the force curves for $F_\gamma^x$ do not shift in a single direction. For example, in Fig.\,\ref{fig:ForceCurves}b, the maximum horizontal force $\max(F_\gamma^x)$ shift upward from $\theta_s=50^\circ$ to $110^\circ$ and downward from $110^\circ$ to $150^\circ$. \textcolor{black}{The reason why the maximum horizontal force does not increase monotonically can be understood by analyzing the orientation of the capillary force (shown by the angle $\alpha$ in Fig.\,\ref{fig:CapillaryBridge}b). The horizontal force is largest when the capillary force vector points purely along $x$ (\textit{i.e.,} when $\alpha=0$). For the case shown in Fig.\,\ref{fig:ForceCurves}b with $\theta_p=50^\circ$, $\alpha$ is closest to zero when $\theta_s=110^\circ$. Therefore, any deviation away from $\theta_s=110^\circ$ causes $\alpha$ to deviate away from zero, which corresponds to the the capillary force being distributed in both the horizontal and vertical direction, rather than purely in the horizontal direction.}

Although the surface wettability influences the maximum horizontal and vertical capillary force components (Fig.\,\ref{fig:ForceCurves}), it does not affect the maximum net capillary force, $F_\gamma^\mathrm{max}=\max\left[\sqrt{(F_\gamma^x)^2+(F_\gamma^z)^2}\right]$. Indeed, plotting $F_\gamma^\mathrm{max}$ for various $\theta_p$ and $\theta_s$ (Fig.\,\ref{fig:ForceCurves}c) reveals that $F_\gamma^\mathrm{max}$ depends mainly on $\theta_p$ and barely on $\theta_s$. In all cases, the maximum capillary force can be described using $F_\gamma^\mathrm{max}= 2\pi R_p \gamma \cos^2(\theta_p/2)$ ~\cite{scheludko_measurement_1975}. Remarkably, this expression provides an excellent fit to the data, even though it was originally derived to predict the detachment force for a particle crossing a liquid-air interface perpendicularly. The excellent fit suggests that the wettability of the underlying solid surface does not significantly influence the shape of the liquid-air-particle contact line during detachment.

However, surface wettability is still important because it influences the orientation of the capillary force. Increasing $\theta_s$ rotates the capillary force vector upward, which can cause it to change sign. While the magnitude of the capillary force depends only on $\theta_p$, its orientation depends on both $\theta_p$ and $\theta_s$, with the angle $\alpha$ increasing with both $\theta_p$ and $\theta_s$ (Fig.\,\ref{fig:ForceCurves}d).

\subsection{Phase diagram for outcomes of drop-particle interaction} 

 \begin{figure*}[t]
     \centering
     \includegraphics{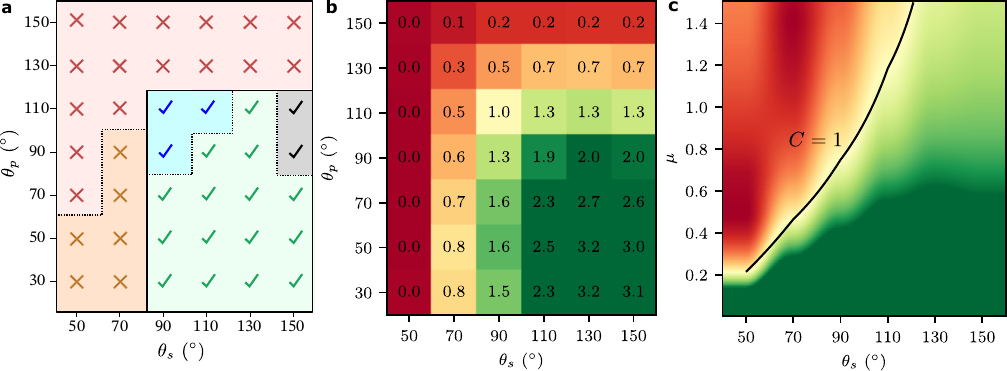}
     \caption{\textbf{Phase diagrams mapping drop--particle interaction scenarios.} (a) Phase diagram showing different collision scenarios as a function of $\theta_s$ and $\theta_p$ for $\mu=0.6$ and $\mu_r=0.3$. All the simulations in this phase diagram correspond to the capillary bridge geometry (Fig.\,\ref{fig:CapillaryBridge}a). The same force was applied to move the drop in all cases. Ticks and crosses mean that the particle was captured or deposited, respectively. Green region: particle entered the front of the drop and remained attached to the rear (Fig.\,\ref{fig:IntroFigure}d). Blue region: particle went around the drop perimeter and remained attached at the rear (Fig.\,\ref{fig:IntroFigure}e). Gray region: \textcolor{black}{particle travelled under the drop and remained attached at the rear of the drop} (Fig.\,\ref{fig:IntroFigure}f and Supplementary Fig.\,S3) Red region: \textcolor{black}{the particle went around the drop and then detached} (\textit{i.e.,} redeposited, Fig.\,\ref{fig:IntroFigure}h). Orange region: \textcolor{black}{particle left in a water film behind the drop} (Fig.\,\ref{fig:IntroFigure}i). For higher $\mu$, an additional outcome is possible where the particle enters and exits the drop, as shown in Fig.\,\ref{fig:IntroFigure}g (top row). (b) Predictions of $C$ (numbers written in the squares) calculated using Eq.\,\ref{Eq. Capture threshold} {\color{black} with $\mu=0.6$.} Supplementary Fig.\,S6 shows predictions for other values of $\mu$. (c) Predicted $C$ as a function of $\mu$ and $\theta_s$ for $\theta_p=90^\circ$. Black line: contour where $C=1$.}
     \label{fig:PhaseDiagram}
 \end{figure*}

To understand how and when changes in the surface and particle wettability lead to the variety of outcomes observed in Fig.\,\ref{fig:IntroFigure}(f,g), we systematically mapped out the outcomes in a phase diagram (Fig.\,\ref{fig:PhaseDiagram}a). In the limit where the friction force is negligible ($\mu\rightarrow 0$), the drop is expected to always capture the particle since there is no resistance to motion. In contrast, when the friction force is significantly greater than the capillary force ($\mu\rightarrow \infty$), the drop is expected to never capture the particle because the capillary force cannot overcome the friction force. Figure\,\ref{fig:PhaseDiagram}a corresponds to the case where the friction force and the capillary force have the same order of magnitude, with $\mu F_p/(\gamma R_p)\sim \mathcal{O}(1)$. We focus on the regime where the friction and capillary forces are comparable because it corresponds to the scenario we observed experimentally with a hydrophobic particle on a hydrophobic surface, as seen in Fig.\,\ref{fig:IntroFigure}d when $\theta_p\approx\theta_s\approx90^\circ$.

Varying $\theta_p$ and $\theta_s$ while keeping everything else constant gave rise to {\color{black}five out of the six collision} scenarios (shown in Fig.\,\ref{fig:IntroFigure}d--f,h,i and Movies 2-4, 6, 7). {\color{black} The sixth scenario (Fig.\,\ref{fig:IntroFigure}g) is obtained when the friction coefficient or the force applied to the drop is increased.} In the phase diagram shown in Figure\,\ref{fig:PhaseDiagram}a, {\color{black}the main scenario that leads to successful capture (green region) corresponds to when the particle enters the drop and remains attached at the rear (Fig.\,\ref{fig:IntroFigure}d). The other two successful capture scenarios (blue and grey regions) occupy a smaller part of the phase diagram. The blue region, corresponding to when the particle goes around the drop and remains attached (Fig.\,\ref{fig:IntroFigure}e), is obtained when both the particle and the surface were neutrally wetting ($\theta_p\approx\theta_s\approx 90^\circ$). The grey region, corresponding to when the particle travels under the drop and remains attached (Fig.\,\ref{fig:IntroFigure}f), is only seen with hydrophobic particles on superhydrophobic surfaces. Despite being the most widely used set of contact angles in studies investigating particle removal, this scenario only occupies a small part of the phase diagram. When the particle is deposited back onto the surface, there are two possible scenarios, shown by the red and orange regions. The red region corresponds to when the particle goes around the front half of the drop before detaching (Fig.\,\ref{fig:IntroFigure}h). This scenario occurs when the horizontal capillary force is large enough to prevent the particle from crossing the front drop-air interface, but insufficient to pull the particle. The orange region corresponds to when the particle was deposited at the rear of the drop in a liquid film (Fig.\,\ref{fig:IntroFigure}i). This scenario typically occurs when the surface is hydrophilic such that the drop elongates to form a film.}

Varying the speed of the drop can change the phase diagram. All the simulations in the phase diagram were obtained by applying a constant force to move the drop, leading to drop speeds that are within the same order of magnitude throughout the phase diagram. In Supplementary Fig.\,S5, we show that reducing the applied force at a fixed $\theta_s=70^\circ$ shifts the boundary between film formation and capture (orange/green boundary) to the left, thus extending the range of $\theta_s$ over which the particle was captured. Capture is more likely when the applied force is lower because reducing the applied force reduces the speed of the drop. A reduction in the drop speed increases the dynamic contact angle at the receding side of the drop~\cite{snoeijer_moving_2013} and makes it less likely for the drop to leave a liquid film behind. Therefore, reducing the speed effectively increases the contact angle $\theta_s$ at the rear of the drop, causing $\alpha$ to increase, thus enhancing the capture parameter (Fig.\,\ref{fig:ForceCurves}d). To maximise the likelihood of capillary-driven capture of particles, the applied force on the drop should be as small as possible while remaining sufficiently large for the drop to overcome friction between the particle and the surface.

Mapping out phase diagrams such as the one shown in Fig.\,\ref{fig:PhaseDiagram}a is computationally demanding because the simulations have to be carried out in three dimensions to allow the particles to move around the drop perimeter. Therefore, it is valuable to be able to predict the phase diagram for different $\theta_p$, $\theta_s$, $\mu$, and $F_p$ without having to perform full simulations. To derive a predictive criterion, we rearrange Eq.\,\ref{Eq. Capture inequality} by grouping all terms into a single dimensionless number, $(F_\gamma^x+\mu F_\gamma^z)/(\mu F_p)$. The maximum value taken by this dimensionless parameter,
\begin{equation}
C=\frac{\max(F_\gamma^x+\mu F_\gamma^z)}{\mu F_p},
\label{Eq. Capture threshold}
\end{equation}
indicates whether the capillary force can overcome the friction force. Since the capillary force drives particle capture, we call $C$ the capillary capture parameter. The particle is expected to be captured when $C>1$. From the force curves for $F_\gamma^x$ and $F_\gamma^z$ (shown in Fig.~\ref{fig:ForceCurves}a,b and in Supplementary Fig.\,S2), we computed $C$ for various $\theta_p$ and $\theta_s$ and obtained an excellent prediction for the phase diagram (Fig.\,\ref{fig:PhaseDiagram}b). 

The capillary capture parameter (Eq.\,\ref{Eq. Capture threshold}) makes it possible to efficiently generate predicted phase diagrams for a wide range of contact angles, friction coefficients, and particle weights/densities. For example, in Supplementary Fig.\,S6, we show predicted phase diagrams for four different friction coefficients (between $\mu=0.01$ and $\mu=1.5$), highlighting that the lower the friction coefficient, the greater the capillary capture parameter for any given contact angle. In Supplementary Fig.\,S7, we show predicted phase diagrams for different particle densities, highlighting that reducing the vertical force $F_p$ on the particle also increases the capillary capture parameter. In experiments, inhomogeneities on the particle or on the surface may still cause the particle to detach for values above, but close to, 1. Therefore, $C$ should be as high as possible to ensure successful capture. 

The capillary capture parameter allows us to understand a range of different scenarios, including the effects of varying the cleaning liquid, the surface chemistry of the surface and particle, and the bulk material of the particle. Using different liquids changes both $\theta_s$ and $\theta_p$. Thus, plotting $C$ as a function of $\theta_s$ and $\theta_p$ (as shown in Fig.\,\ref{fig:PhaseDiagram}b) corresponds to using different liquids while maintaining the same material for the particle and the surface. In contrast, varying the surface chemistry of the surface influences both $\theta_s$ and $\mu$, in which case it is helpful to plot $C$ against $\theta_s$ and $\mu$ instead (Fig.\,\ref{fig:PhaseDiagram}c). Plotting the capillary capture parameter in this form highlights that on superhydrophobic surfaces ($\theta_p\gtrapprox140^\circ$), particles with a wider range of friction coefficients can be captured (Fig.\,\ref{fig:PhaseDiagram}c).

\begin{figure*}
     \centering
     \includegraphics[width=0.85\linewidth]{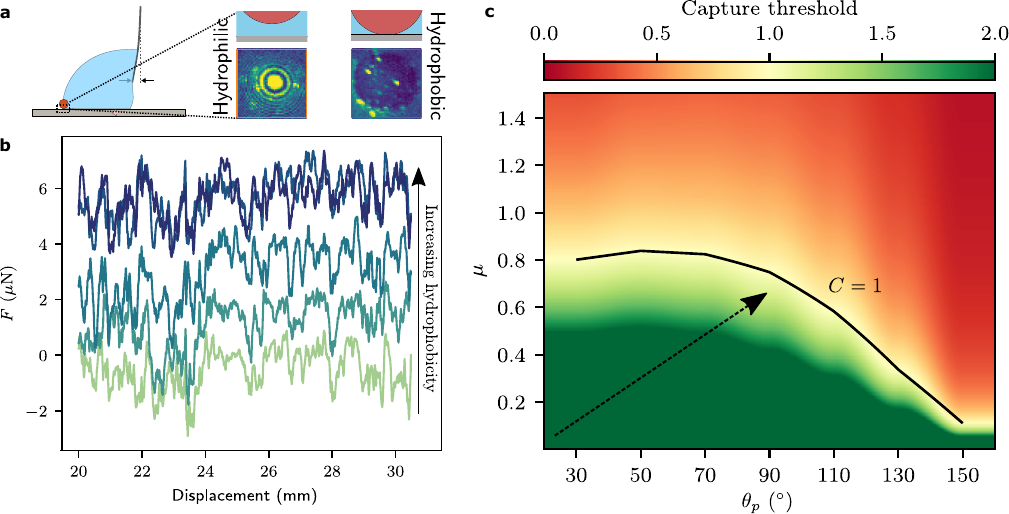}
     \caption{\textbf{Variation of particle friction as particle becomes more hydrophobic.} (a) Interference fringes observed when imaging the reflection of the laser beam from below using an inverted confocal microscope confirm the presence of a water film under the hydrophilic particle. No interference fringes were present below the hydrophobic particle, indicating the absence of a water film. Speed: 100\,$\mu$m/s. (b) Force on the particle when it was pulled by the drop for 5 successive runs of the same experiment. The force was measured from the cantilever deflection. Each force curve corresponds to a different run using the same particle. The bottommost curve (light green) corresponds to the first run and the topmost curve (dark blue) corresponds to the last run.  Initially, the particle was hydrophilic ($\theta_p\approx 20^\circ$). The force increased after each run up to the fourth run, indicating that the particle friction increases due to the accumulation of free PDMS chains. The curves for the last two runs almost overlap, indicating that the particle became saturated with PDMS. (c) Dependence of the capture parameter as a function of $\mu$ and $\theta_p$ for $\theta_s=90^\circ$. In this diagram, $C$ was calculated using Eq.\,\ref{Eq. Capture threshold} and the capillary force curves in Fig.\,\ref{fig:ForceCurves}b and Supplementary Fig.\,S2. The dashed arrow indicates the general direction in which we expect $\mu$ and $\theta_p$ to increase as the contact time between the particle and the surface increases in (b).}
          \label{fig:Interference}
 \end{figure*}

To illustrate the importance of the friction coefficient, we return to the experiments shown in Fig.\,\ref{fig:IntroFigure} {\color{black}on the hydrophobic PDMS surface}. Imaging the contact area between the particle and the surface using interference microscopy revealed the presence of a water film under the hydrophilic particle (Fig.\,\ref{fig:Interference}a). {\color{black}A potential explanation for the water film is that in water, both the glass particle and the PDMS surface acquire an effective negative charge~\cite{gu_potential_2000, liu_detecting_2020}. An electric double layer forms around both the PDMS surface and the glass particle, leading to a repulsive force. The range of this repulsive electrostatic force is of the order of the Debye length, which is of the order of hundreds of nanometres for pure water~\cite{israelachvili_intermolecular_2011}, leading to visible interference fringes. We note that Van der Waals forces alone cannot explain the presence of a water film because they are attractive between glass and PDMS (see Supplementary Section III).} The water film lubricates the contact between the hydrophilic particle and the surface, reducing the coefficient of friction. Consequently, the force exerted by the drop to pull the hydrophilic particle was significantly lower than the force to pull a hydrophobic particle, as indicated, respectively, by the bottommost and topmost force curve in Fig.\,\ref{fig:Interference}b.

To understand how the forces evolve between the hydrophilic and hydrophobic limits, we varied the wettability of the particle. Pristine hydrophilic particles accumulate uncrosslinked PDMS chains~\cite{wong_adaptive_2020,naga_how_2021,hourlier-fargette_role_2017} when in contact with the PDMS surface. This causes the hydrophobicity of the particle to increase as the particle rolls on the surface. Therefore, by repeating the force measurements with the same particle along the same surface trajectory, we systematically probed the effect of increasing particle hydrophobicity while eliminating variations due to surface heterogeneities. Successive measurements produced force curves that shifted progressively upward until the fourth measurement, beyond which no further change was observed, indicating saturation of the particle surface with PDMS chains.

The identical fluctuation patterns observed in all force curves confirm that the measurements were performed over the same surface trajectory. The consequences of the increasing particle hydrophobicity are captured by the phase diagram of the capture parameter $C$ as a function of the friction coefficient $\mu$ and particle contact angle $\theta_p$ (Fig.\,\ref{fig:Interference}c). Experimentally, the increase in hydrophobicity caused by the accumulation of PDMS chains on the particle leads to simultaneous increases in $\theta_p$ and $\mu$, the latter arising from thinning and eventual rupture of the lubricating water film beneath the particle. In the phase diagram, this evolution corresponds to a trajectory from the lower left toward the $C=1$ boundary, as indicated by the dashed black line in Fig.\,\ref{fig:Interference}c (here a straight line is sketched for simplicity).

\section{Conclusions}

{\color{black}We presented phase diagrams for particle capture and introduced a dimensionless capillary capture parameter (Eq.\,\ref{Eq. Capture threshold}) to predict particle removal. The capillary capture parameter accounts for the interplay between capillary and friction forces on the particle. While the magnitude of the capillary force is largely independent of the surface wettability, its orientation is sensitive to both the particle wettability and the surface wettability. The capillary force can either point downward, thus increasing friction and hindering particle removal, or upward, thus reducing friction and promoting particle removal. Consequently, maximizing surface hydrophobicity is not always better for particle removal. For example, in Figure\,\ref{fig:PhaseDiagram}b, the value of the capillary capture parameter is higher when $\theta_s=130^\circ$ than when $\theta_s=150^\circ$. This effect is because on a highly superhydrophobic surface, the capillary force vector is oriented more towards the vertical direction, causing a reduction in the horizontal component that acts to pull the particle along the surface. The optimal surface contact angle for an easy-to-clean surface corresponds to the value that maximizes the capillary capture parameter and is a function of both the friction coefficient and the normal contact force between the particle and the surface.

When designing easy-to-clean surfaces, it is crucial to minimize particle--surface friction since the capillary capture parameter in Eq.\,\ref{Eq. Capture threshold} is highly sensitive to the $\mu$. It is also useful to reduce contact angle hysteresis between drops and surfaces. The key consideration is that the receding drop contact angle must be sufficiently large to ensure the drop maintains its shape and does not stretch to leave a water film behind. The best strategy is to ensure that the particle--surface material combination lies comfortably on the correct side of the phase boundary rather than aiming to achieve the maximum possible value for the capillary capture parameter. This strategy offers more flexibility to help balance other aspects, such as durability or ease of fabrication, that are also crucial to ensure the long-term performance of easy-to-clean surfaces. For instance, our findings highlight that particle removal is possible on hydrophobic surfaces, without the need for superhydrophobicity. Since hydrophobic surfaces can be achieved without fragile nano/micro surface textures required for superhydrophobic surfaces, easy-to-clean surfaces can in principle be achieved using more durable hydrophobic coatings~\cite{ma_ultra-thin_2021,karakaya_silica_2026} that do not require surface textures.}

{\color{black} The removal of hydrophilic and hydrophobic particles is different on PDMS surfaces because hydrophilic particles lose direct contact with the surface and aquaplane on a water film. The water film significantly reduces the friction force between the particle and the surface. Therefore, a hydrophilic particle can be removed much more easily for two reasons: it experiences a higher capillary force from the drop and a lower friction when moving on the surface. When predicting particle removal, it is crucial to quantify the adhesion force and the friction coefficient in the relevant cleaning liquid because the liquid can modify the interactions between the particle and the surface. For example, while a glass particle experiences an attractive van der Waals force with PDMS in air, in water, an electric double layer is expected to form, leading to an electrostatic repulsion that supports aquaplaning.  Alternatively, since measuring the friction coefficient can be challenging, we can deduce it by reversing how the capillary capture parameter is used. This would involve first performing experiments to map out the phase diagram and then inferring the friction coefficient by identifying the value that matches the predicted phase diagram.}

For practical applications, it is also interesting to consider different types of particles, which can differ both in terms of their bulk material and their surface chemistry. The bulk material influences the particle density and thus the gravitational force acting on the particle. The surface chemistry influences the van der Waals and electrostatic forces between the particle and the surface, as well as the contact angle $\theta_p$ and friction coefficient $\mu$. When the upward component of the capillary force exceeds the downward force $F_p$ on the particle, the particle gets lifted off the surface (for example, see Supplementary Fig.\,S7). In this case, the particle is very easily captured because the effective friction force decreases significantly due to the loss of contact. For hydrophobic particles on hydrophilic surfaces, the snap-in event when the particle first encounters the liquid-air interface can also lead to a significant upward capillary force (Fig.\,\ref{fig:ForceCurves}a). One difference between the van der Waals and gravitational forces is that the van der Waals force is short-ranged and becomes inactive once the particle loses contact with the surface. Therefore, once contact is lost, the particle can easily get lifted off the surface, following the flow inside the drop. In contrast, the gravitational force persists even when contact is lost. Therefore, larger particles will only get lifted off when the viscous force due to the flow of liquid in the drop exceeds the gravitational force.

This paper lays the groundwork for predicting the removal of particles by drops, opening opportunities to tackle even more complex questions related to interfacial cleaning processes. For example, there remain major challenges to understand the effect of particle shape, to explore how the removal mechanism evolves when multiple particles are present, and to understand the effect of surface heterogeneities. Particles with irregular shapes may be captured in some orientations, but not in others. When multiple particles are present, friction and adhesion between adjacent particles will need to be considered since they will affect the particles' motion. On heterogeneous surfaces with physical textures or chemical patterns, particles may get captured in some regions, but not in others. 

Besides the direct relevance to interfacial cleaning processes, our findings are also useful for other applications. For example, the capillary assisted particle assembly technique exploits the interplay between capillary and friction forces to deposit colloidal particles in controlled locations on a surface for applications in optics, electronics, and biomaterials~\cite{ni_capillary_2018, liu_capillary_2018}. Our phase diagrams can also be used to understand the limits of capillary assembly. For cleaning applications, it is beneficial that the capillary capture parameter is as large as possible, whereas for capillary assembly, the capillary capture parameter must be neither too large nor too small to ensure particles get deposited, but only in desired locations.

\section{Materials and methods}
\label{Section: Methods}

\subsection{Numerical method}

The numerical method that we used for the simulations method combines the lattice Boltzmann method (LBM) to model the dynamics of the two fluids (liquid and air) and the discrete element method (DEM) to model contact forces (normal reaction, sliding friction, rolling friction) between the particles and the flat surface. The method is described in detail in~\cite{naga_modeling_2025}. The method explicitly accounts for the key interactions required for studying particle removal, including capillary forces between particles and liquid-fluid interfaces, friction forces between the particle and the surface, and hydrodynamic forces between the fluids and particles.

The tunable parameters of the numerical model include the particle contact angle $\theta_p$, the surface contact angle $\theta_s$, the coefficient of friction $\mu$ (both for sliding and rolling friction), the force acting on the particle, and the force acting on the drop. These parameters can be tuned independently, allowing us to systematically study the effects of each parameter.

In this paper, we kept the surface tension and viscosities of the fluid constant to focus on investigating the effect of varying the contact angles and friction force. The surface tension was set to $\gamma=72\,$mN/s ($\gamma=0.01$ in simulation units, see Supplementary Table S2 for a summary of parameters and explanation on unit conversion). The dynamic viscosity of the drop was set to $\eta_d=1$\,mPa\,s ($\eta_d=0.0274$ in simulation units). The dynamic viscosity of the surrounding air phase was 10 times lower than the dynamic viscosity of the drop. The particle radius was $100\,\mu$m ($10$ lattice units). The downward vertical force acting on the particle was chosen such that the capillary force ($\sim\gamma R_p$) and the friction force ($\sim \mu F_p$) have a similar order of magnitude, $\mu F_p/(\gamma R_p)\approx2$. We varied $\theta_p$ and $\theta_s$ between $30^\circ$ and $150^\circ$.

In Fig.\,\ref{fig:IntroFigure}f,g, we simulated a whole drop to show the drop-particle interaction fully in three-dimensions. For these simulations, the drop was initialized as a hemisphere with radius 500\,$\mu$m (50 lattice units). A force between $94\,\mu$N and $113\,\mu$N was applied to move the drop along $x$, corresponding to a Bond number ($\mathrm{Bo}=f_xR_d^2/\gamma$, where $f_x$ is the force applied to the drop per unit volume) between 1.25 and 1.5. The coefficients of sliding and rolling friction were set to 0.6 and 0.3, respectively, except for the top row of Fig.\,\ref{fig:IntroFigure}g where both coefficients were set to 1.0. The particle radius was 100\,$\mu$m (10 lattice units). The initial positions of the drop and the particle were offset by 20\,$\mu$m (2 lattice units) along the $y$ axis to allow the particle to break the symmetry along $y$ and move around the drop when it does not fully enter at the front. In all the simulations in this paper, we imposed periodic boundary conditions along $x$ (the direction of motion) and $y$.

The experiments and simulations shown in Fig.\,\ref{fig:IntroFigure} highlight that only the bottom part of the drop influences the interaction. Therefore, beyond Fig.\,\ref{fig:IntroFigure} in the paper, we switch to a capillary bridge geometry where a cylindrical drop is sandwiched between two parallel solid surfaces. Besides saving computational cost, the capillary bridge geometry has the advantage of controlling the speed of the drop. Without the top surface, the drop speed would differ significantly when the contact angle on the bottom surface ($\theta_s$) is varied while keeping the force applied to the drop constant. Since the top surface does not significantly influence the shape of the bottom of the drop (which is relevant for the drop-particle interaction), we used the capillary bridge geometry to generate data for all the force curves and phase diagrams shown in the paper.

To obtain the force curves shown in Fig.\,\ref{fig:ForceCurves}, we fixed the particle while moving the capillary bridge to the right as shown in Fig.\,\ref{fig:CapillaryBridge}. For this set of simulations, the domain size was $2.4\,\mathrm{mm}\times0.6\,\mathrm{mm}\times0.6\,\mathrm{mm}$ ($400\times60\times60$ lattice units), along $x$, $y$, and $z$, respectively. The length of the capillary bridge along $x$ was 1\,mm (100 lattice units), and its width along $y$ was 0.6\,mm (60 lattice units). The particle radius was 100\,$\mu$m (10 lattice units). We applied a force of 48\,$\mu$N along $x$ to move the capillary bridge. For these simulations, the coefficient of friction and the gravitational force on the particle are not relevant because the particle is fixed in position. The contact angle $\theta_s$ between the liquid and the bottom surface was varied between $30^\circ$ and $150^\circ$. On the top plate, the contact angle was set to ($180^\circ-\theta_s$) such that the rear liquid-air interface is initially straight to mimic what would be observed when a large drop interacts with a small particle, where the drop radius is much larger than the particle radius.

For the phase diagram presented in Fig.\,\ref{fig:PhaseDiagram}a, the domain size was $2.4\,\mathrm{mm}\times1.8\,\mathrm{mm}\times0.6\,\mathrm{mm}$ ($400\times180\times60$ lattice units), along $x$, $y$, and $z$, respectively. The contact angle at the top plate was fixed to $90^\circ$ while the contact angle at the bottom surface was varied from $50^\circ$ to $150^\circ$. The drop was initialized as a cylinder of radius $R_d=0.5$\,mm (50 lattice units), with the long axis along $z$. Along the $x$ direction, the center of the particle was positioned 0.6\,mm in front of the advancing side of the drop. Along the $y$ direction, the center of the particle was positioned 20\,$\mu$m (2 lattice units) to the right of the drop. This offset is to allow the particle to break the symmetry along $y$ and move around the drop when it does not fully enter at the front. The coefficients of sliding and rolling friction were set to 0.6 and 0.3, respectively. A horizontal force of $158\,\mu$N was applied to the drop along $x$. This force was chosen so that the force per unit volume on the drop was the same as in the simulations in Fig.\,\ref{fig:IntroFigure} ($f_x=359\,\mathrm{N/m^3}$ or $5\times10^{-6}$ in simulation units). This force corresponds to a Bond number of $\mathrm{Bo}=f_xR_d^2/\gamma=1.25$.

The computational cost to generate each simulation data point in the phase diagrams presented in this paper is $\approx500$ core hours on the Cirrus UK National Tier-2 HPC Service. Generating the full phase diagram involves a substantial computational cost of tens of thousands of core hours. The predictions provided by the capture parameter proposed in this paper therefore make it possible to efficiently predict the outcomes as well as to explore the effects of varying a wide range of relevant parameters ($\theta_p$, $\theta_s$, $\mu$, $\rho_p$) without having to perform full simulations. 

\subsection{Experimental details}

The experiments were carried out using an inverted laser scanning confocal microscope (Leica TCS SP8) coupled with a metallic blade to measure friction forces. The setup was introduced in detail in~\cite{naga_how_2021}. Briefly, a flexible metal blade (stainless steel, dimensions 60\,mm$\times$4.6\,mm$\times$0.07\,mm, spring constant $k=0.329$\,N/m) was mounted above the objective lens to fix the drop in position while moving the underlying substrate at a well-defined speed (100\,$\mu$m/s throughout the paper). The blade was hydrophobic with a contact angle $\approx 90^\circ$ with water.

Measuring the deflection $\Delta x$ of the blade using the reflected laser from the microscope also allowed us to obtain the total horizontal force acting on the drop and the particle in the direction of motion using Hooke's law, $F_\mathrm{tot}=k \Delta x$~\cite{pilat_dynamic_2012,gao_how_2018}. To obtain the force on the particle, we subtracted the force curve for a water drop only (without the particle) from the force curves corresponding to when the drop interacts with the particle. Both force curves involved in this subtraction were measured along the same trajectory on the surface with the same velocity.
 
For the confocal microscopy experiments, we used small water drops (volume 2\,$\mu$L) to ensure that the entire drop footprint fits within the field of view of the microscope. Fluorescent dye (ATTO 488, concentration $10^{-3}$\,mg/mL) was added to the water to enable fluorescent imaging using confocal microscopy. The hydrophobic polydimethylsiloxane (PDMS) surfaces used in the confocal microscope experiments were prepared by spin coating a Sylgard 184 mixture with 10 parts base and 1 part crosslinker (by weight) onto a cleaned microscope coverslip. The advancing and receding contact angles of water on PDMS surfaces are $120^\circ$ and $80^\circ$, respectively~\cite{wong_adaptive_2020}.

{\color{black}For side-view imaging, we also performed experiments using a goniometer (Krüss DSA100) coupled either with the same metallic blade mentioned before or a glass capillary (length of the free end 49\,mm, width  0.5\,mm, spring constant 0.121\,N/m). We conducted experiments varying the wettability of both the surface and the particle, ranging from hydrophilic to superhydrophobic.

To prepare hydrophilic surfaces, microscope glass coverslips were plasma cleaned for 30\,s at 300\,W. The resulting contact angle was $<30^\circ$. For hydrophobic surfaces, in addition to the PDMS surface described above, we prepared polystyrene-coated microscope glass coverslips. The advancing and receding contact angles of the polystyrene surfaces were $100^\circ$ and $80^\circ$, respectively. To prepare the polystyrene coating, a 4 wt\% solution of polystyrene (molecular weight of 214 000) in toluene was spin coated onto a cleaned microscope coverslip. Finally, the coating was cured at 120$^\circ$C for 24\,h. As superhydrophobic surface, we coated glass slides with silicone nanofilaments~\cite{geyer_when_2020}. The procedure was as follows: 1.4\,mL of trichloromethylsilane (99\%, Sigma Aldrich) was added to a Teflon reaction chamber containing 350\,mL of toluene (Sigma Aldrich) with water content of 155\,ppm. The solution was stirred for 60 seconds, then the microscope coverslip was immersed and the reaction chamber was sealed. After 3\,h the coverslip was rinsed with n-hexane (95\%, Fischer chemical) and dried under a nitrogen stream.

To prepare hydrophilic particles, we plasma-treated glass particles three times for 30\,s at 300\,W. Freshly prepared hydrophilic particles had a contact angle of $\approx 20^\circ$, as measured by imaging the water meniscus between the particle and a water drop using confocal microscopy. When in contact with a PDMS surface, the hydrophilic particle accumulated uncrosslinked PDMS chains from the surface. Therefore, using the same initially hydrophilic particle for several consecutive measurements on PDMS surfaces gradually increased the contact angle of the particle from $\approx20^\circ$ until it saturated to $\approx 110^\circ$. This effect allowed us to vary the particle wettability to obtain particles with intermediate wettabilities used for experiments on the other surfaces. To obtain even higher contact angles and achieve superhydrophobic particles, we coated glass particles twice with a commercial superhydrophobic coating (SOFT99 Glaco Mirror Coat). This procedure gives particle contact angles of $\approx 145^\circ$. Table\,\ref{Table: Experimental details} provides a summary of all the materials used and lists experimental details on the drop volume and velocity for each particle--surface combination tested.}

\begin{table}[h]
\centering
\caption{\textbf{Summary of materials and parameters used in the experiments.} Please note the contact angles listed here are approximate only serve as a guide to interpret the results. These values lie between the advancing and receding contact angles.}
\label{Table: Experimental details}
\begin{tabular}{@{}lllcllccc@{}}
\hline
Instrument & Blade & Surface & $ \theta_s $ ($ {}^\circ $) & $2R_p$ ($ \mu\mathrm{m} $) & Particle & $ \theta_p $ ($ {}^\circ $) & $ \Omega $ ($ \mu\mathrm{L} $) & $ V $ (mm/s) \\
\hline
Goniometer & Metal blade & Clean glass & $ <30 $ & 355--420 & No treatment & 50 & 20 & 1 \\
Goniometer & Metal blade & Clean glass & $ <30 $ & 210--250 & PDMS coated & 105 & 20 & 1 \\
Goniometer & Metal blade & Clean glass & $ <30 $ & 210--250 & Glaco coated & 145 & 20 & 0.5 \\
Confocal & Metal blade & PDMS & 100 & 210--250 & Plasma cleaned & 20 & 2 & 0.1 \\
Confocal & Metal blade & PDMS & 100 & 210--250 & PDMS coated & 105 & 2 & 0.1 \\
Goniometer & Metal blade & PS & 90 & 210--250 & Glaco coated & 145 & 10 & 0.5 \\
Goniometer & Metal blade & Nanofilaments & 150 & 210--250 & PDMS coated & 105 & 10 & 1.2 \\
Goniometer & Glass capillary & Nanofilaments & 150 & 210--250 & Glaco coated & 145 & 10 & 0.5 \\
\hline
\end{tabular}%

\end{table}

To determine whether a particle aquaplaned on a water film, we checked for the presence of interference fringes underneath the particle. When a water film is present, the laser beam from the confocal microscope gets reflected at the surface-water interface and the water-particle interface. These reflected signals interfere to create interference fringes. For this procedure, we used 458\,nm and 633\,nm lasers, focusing on the three-phase contact line following the procedure described in~\cite{daniel_oleoplaning_2017}. The presence of interference fringes indicates that a water film is present between the particle and the surface, whereas their absence indicates that the particle is in direct contact with the surface.

\bibliography{apssamp}

@article{gu_potential_2000,
	title = {The $\zeta$-{Potential} of {Glass} {Surface} in {Contact} with {Aqueous} {Solutions}},
	volume = {226},
	doi = {10.1006/jcis.2000.6827},

	number = {2},
	journal = {J. Colloid Interf. Sci.},
	author = {Gu, Yongan and Li, Dongqing},
	year = {2000},

	pages = {328--339},
}

@article{liu_detecting_2020,
	title = {Detecting zeta potential of polydimethylsiloxane ({PDMS}) in electrolyte solutions with atomic force microscope},
	volume = {578},
	doi = {10.1016/j.jcis.2020.05.061},
	journal = {J. Colloid Interf. Sci.},
	author = {Liu, Zhijian and Song, Yongxin and Li, Dongqing},
	year = {2020},
	pages = {116--123},
}

@article{geyer_when_2020,
    title = {When and how self-cleaning of superhydrophobic surfaces works},
    volume = {6},
    doi = {10.1126/sciadv.aaw9727},
    number = {3},
    journal = {Sci. Adv.},
    author = {Geyer, Florian and D’Acunzi, Maria and Sharifi-Aghili, Azadeh and Saal, Alexander and Gao, Nan and Kaltbeitzel, Anke and Sloot, Tim-Frederik and Berger, Rüdiger and Butt, Hans-Jürgen and Vollmer, Doris},
    year = {2020},
    pages = {eaaw9727},
}

@article{naga_modeling_2025,
	title = {Modelling droplet–particle interactions on solid surfaces by coupling the lattice {Boltzmann} and discrete element methods},
	volume = {1040},
	doi = {10.1017/jfm.2026.11865},
	journal = {J. Fluid Mech.},
	author = {Naga, Abhinav and Zhang, Xitong and Yang, Junyu and Kusumaatmaja, Halim},
	year = {2026},
	pages = {A21},
}

@article{naga_how_2021,
    title = {How a water drop removes a particle from a hydrophobic surface},
    volume = {17},
    doi = {10.1039/D0SM01925A},
    number = {7},
    journal = {Soft Matter},
    author = {Naga, Abhinav and Kaltbeitzel, Anke and Wong, William S. Y. and Hauer, Lukas and Butt, Hans-Jürgen and Vollmer, Doris},
    year = {2021},
    pages = {1746--1755},
}

@book{butt_surface_2018,
    edition = {2},
    title = {Surface and {interfacial} {forces}},
    isbn = {978-3-527-80436-8},
    publisher = {Wiley-VCH},
    author = {Butt, Hans-Jürgen and Kappl, Michael},
    year = {2018},
}

@article{scheludko_measurement_1975,
    title = {Measurement of surface tension by pulling a sphere from a liquid},
    volume = {253},
    doi = {10.1007/BF01382159},
    number = {5},
    journal = {Colloid Polym. Sci.},
    author = {Scheludko, A. D. and Nikolov, D.},
    year = {1975},
    pages = {396--403},
}

@article{heckenthaler_self-cleaning_2019,
    title = {Self-Cleaning Mechanism: Why Nanotexture and Hydrophobicity Matter},
    volume = {35},
    doi = {10.1021/acs.langmuir.9b01874},
    number = {48},
    journal = {Langmuir},
    author = {Heckenthaler, Tabea and Sadhujan, Sumesh and Morgenstern, Yakov and Natarajan, Prakash and Bashouti, Muhammad and Kaufman, Yair},
    year = {2019},
    pages = {15526--15534},
}

@article{wang_self-cleaning_2025,
    title = {Self-cleaning properties of superhydrophobic surface: Effect of particle wettability},
    volume = {452},
    shorttitle = {Self-cleaning properties of superhydrophobic surface},
    doi = {10.1016/j.powtec.2024.120536},
    journal = {Powder Technol.},
    author = {Wang, Jiu-Si and Li, Chao-Sheng and Cai, Rong-Rong and Zhang, Li-Zhi},
    year = {2025},
    pages = {120536},
}

@article{neinhuis_characterization_1997,
    title = {Characterization and distribution of water-repellent, self-cleaning plant surfaces},
    volume = {79},
    doi = {10.1006/anbo.1997.0400},
    number = {6},
    journal = {Ann. Bot.},
    author = {Neinhuis, C. and Barthlott, W.},
    year = {1997},
    pages = {667--677},
}

@article{hassan_self-cleaning_2019,
    title = {Self-cleaning of a hydrophobic surface by a rolling water droplet},
    volume = {9},
    doi = {10.1038/s41598-019-42318-3},
    number = {1},
    journal = {Sci. Rep.},
    author = {Hassan, Ghassan and Yilbas, Bekir Sami and Al-Sharafi, Abdullah and Al-Qahtani, Hussain},
    month = apr,
    year = {2019},
    pages = {5744},
}

@article{perumanath_contaminant_2023,
    title = {Contaminant Removal from Nature’s Self-Cleaning Surfaces},
    volume = {23},
    doi = {10.1021/acs.nanolett.3c00257},
    number = {10},
    journal = {Nano Lett.},
    author = {Perumanath, Sreehari and Pillai, Rohit and Borg, Matthew K.},
    year = {2023},
    pages = {4234--4241},
}

@article{yin_dynamic_2018,
    title = {Dynamic Effects on the Mobilization of a Deposited Nanoparticle by a Moving Liquid-Liquid Interface},
    volume = {121},
    doi = {10.1103/PhysRevLett.121.238002},
    number = {23},
    journal = {Phys. Rev. Lett.},
    author = {Yin, Tianya and Shin, Donglee and Frechette, Joelle and Colosqui, Carlos E. and Drazer, German},
    year = {2018},
    pages = {238002},
}

@article{naga_capillary_2021,
    title = {Capillary {torque} on a {particle} {rotating} at an {interface}},
    volume = {37},
    doi = {10.1021/acs.langmuir.1c00851},
    number = {24},
    journal = {Langmuir},
    author = {Naga, Abhinav and Vollmer, Doris and Butt, Hans-Jürgen},
    month = jun,
    year = {2021},
    pages = {7457--7463},
}

@article{marshall_capillary_2014,
    title = {Capillary torque on a rolling particle in the presence of a liquid film at small capillary numbers},
    volume = {108},
    doi = {10.1016/j.ces.2014.01.003},
    journal = {Chem. Eng. Sci.},
    author = {Marshall, Jeffrey S.},
    year = {2014},
    pages = {87--93},
}

@article{lafuma_superhydrophobic_2003,
    title = {Superhydrophobic states},
    volume = {2},
    doi = {10.1038/nmat924},
    number = {7},
    journal = {Nat. Mater.},
    author = {Lafuma, Aurélie and Quéré, David},
    year = {2003},
    pages = {457--460},
}

@article{furstner_wetting_2005,
    title = {Wetting and self-cleaning properties of Artificial Superhydrophobic Surfaces},
    volume = {21},
    doi = {10.1021/la0401011},
    number = {3},
    journal = {Langmuir},
    author = {Fürstner, Reiner and Barthlott, Wilhelm and Neinhuis, Christoph and Walzel, Peter},
    year = {2005},
    pages = {956--961},
}

@article{bhushan_self-cleaning_2009,
    title = {Self-Cleaning Efficiency of Artificial Superhydrophobic Surfaces},
    volume = {25},
    doi = {10.1021/la803860d},
    number = {5},
    journal = {Langmuir},
    author = {Bhushan, Bharat and Jung, Yong Chae and Koch, Kerstin},
    year = {2009},
    pages = {3240--3248},
}

@article{leenaars_particle_1989,
    title = {Particle removal from silicon substrates using surface tension forces},
    number = {44},
    journal = {Philips J. Res.},
    author = {Leenaars, A.F.M. and O'Brien, S.B.G.},
    year = {1989},
    pages = {183--209},
}

@article{gao_frictional_2004,
    title = {Frictional {forces} and {Amontons}' {law}: {from} the {molecular} to the {macroscopic} {scale}},
    volume = {108},
    shorttitle = {Frictional {Forces} and {Amontons}' {Law}},
    doi = {10.1021/jp036362l},
    number = {11},
    journal = {J. Phys. Chem. B},
    author = {Gao, Jianping and Luedtke, W. D. and Gourdon, D. and Ruths, M. and Israelachvili, J. N. and Landman, Uzi},
    month = mar,
    year = {2004},
    pages = {3410--3425},
}

@article{fei_coupled_2023,
    title = {Coupled lattice {Boltzmann} method–discrete element method model for gas–liquid–solid interaction problems},
    volume = {975},
    doi = {10.1017/jfm.2023.822},
    journal = {J. Fluid Mech.},
    author = {Fei, Linlin and Qin, Feifei and Wang, Geng and Huang, Jingwei and Wen, Binghai and Zhao, Jianlin and Luo, Kai H. and Derome, Dominique and Carmeliet, Jan},
    year = {2023},
    pages = {A20},
}

@article{jiang_coupled_2022,
    title = {A coupled {LBM}-{DEM} method for simulating the multiphase fluid-solid interaction problem},
    volume = {454},
    doi = {10.1016/j.jcp.2022.110963},
    journal = {J. Comput. Phys.},
    author = {Jiang, Fei and Liu, Haihu and Chen, Xian and Tsuji, Takeshi},
    year = {2022},
    pages = {110963},
}

@article{blossey_self-cleaning_2003,
    title = {Self-cleaning surfaces — virtual realities},
    volume = {2},
    doi = {10.1038/nmat856},
    number = {5},
    journal = {Nat. Mater.},
    author = {Blossey, Ralf},
    year = {2003},
    pages = {301--306},
}

@article{snoeijer_moving_2013,
    title = {Moving {contact} {lines}: Scales, {regimes}, and {dynamical} {transitions}},
    volume = {45},
    doi = {10.1146/annurev-fluid-011212-140734},
    number = {1},
    urldate = {2021-09-24},
    journal = {Annu. Rev. Fluid Mech.},
    author = {Snoeijer, Jacco H. and Andreotti, Bruno},
    year = {2013},
    pages = {269--292},
}

@article{wong_adaptive_2020,
    title = {Adaptive {Wetting} of {Polydimethylsiloxane}},
    volume = {36},
    doi = {10.1021/acs.langmuir.0c00538},
    number = {26},
    journal = {Langmuir},
    author = {Wong, William S. Y. and Hauer, Lukas and Naga, Abhinav and Kaltbeitzel, Anke and Baumli, Philipp and Berger, Rüdiger and D‘Acunzi, Maria and Vollmer, Doris and Butt, Hans-Jürgen},
    year = {2020},
    pages = {7236--7245},
}

@article{soheili_effect_2023,
    title = {The {effect} of {dust} {deposition} on the {morphology} and {physiology} of {tree} {foliage}},
    volume = {234},
    url = {https://doi.org/10.1007/s11270-023-06349-x},
    doi = {10.1007/s11270-023-06349-x},
    number = {6},
    journal = {Water Air Soil Pollut.},
    author = {Soheili, Forough and Woodward, Stephan and Abdul-Hamid, Hazandy and Naji, Hamid Reza},
    year = {2023},
    pages = {339},
}

@article{beckett_urban_1998,
    title = {Urban woodlands: their role in reducing the effects of particulate pollution},
    volume = {99},
    shorttitle = {Urban woodlands},
    doi = {10.1016/S0269-7491(98)00016-5},
    number = {3},
    journal = {Environ. Pollut.},
    author = {Beckett, K. P. and Freer-Smith, P. H. and Taylor, G.},
    year = {1998},
    pages = {347--360},
}

@article{ilse_techno-economic_2019,
    title = {Techno-{Economic} {Assessment} of {Soiling} {Losses} and {Mitigation} {Strategies} for {Solar} {Power} {Generation}},
    volume = {3},
    doi = {10.1016/j.joule.2019.08.019},
    number = {10},
    journal = {Joule},
    author = {Ilse, Klemens and Micheli, Leonardo and Figgis, Benjamin W. and Lange, Katja and Daßler, David and Hanifi, Hamed and Wolfertstetter, Fabian and Naumann, Volker and Hagendorf, Christian and Gottschalg, Ralph and Bagdahn, Jörg},
    year = {2019},
    pages = {2303--2321},
}

@article{adebiyi_review_2023,
    title = {A review of coarse mineral dust in the {Earth} system},
    volume = {60},
    doi = {10.1016/j.aeolia.2022.100849},
    journal = {Aeolian Res.},
    author = {Adebiyi, Adeyemi and Kok, Jasper F. and Murray, Benjamin J. and Ryder, Claire L. and Stuut, Jan-Berend W. and Kahn, Ralph A. and Knippertz, Peter and Formenti, Paola and Mahowald, Natalie M. and Pérez García-Pando, Carlos and Klose, Martina and Ansmann, Albert and Samset, Bjørn H. and Ito, Akinori and Balkanski, Yves and Di Biagio, Claudia and Romanias, Manolis N. and Huang, Yue and Meng, Jun},
    year = {2023},
    pages = {100849},
}

@article{hourlier-fargette_role_2017,
    title = {Role of uncrosslinked chains in droplets dynamics on silicone elastomers},
    volume = {13},
    doi = {10.1039/C7SM00447H},
    number = {19},
    journal = {Soft Matter},
    author = {Hourlier-Fargette, Aurélie and Antkowiak, Arnaud and Chateauminois, Antoine and Neukirch, Sébastien},
    year = {2017},
    pages = {3484--3491},
}

@article{daniel_oleoplaning_2017,
    title = {Oleoplaning droplets on lubricated surfaces},
    volume = {13},
    doi = {10.1038/nphys4177},
    number = {10},
    journal = {Nat. Phys.},
    author = {Daniel, Dan and Timonen, Jaakko V. I. and Li, Ruoping and Velling, Seneca J. and Aizenberg, Joanna},
    year = {2017},
    pages = {1020--1025},
}

@article{pilat_dynamic_2012,
    title = {Dynamic {Measurement} of the {Force} {Required} to {Move} a {Liquid} {Drop} on a {Solid} {Surface}},
    volume = {28},
    doi = {10.1021/la3041067},
    number = {49},
    journal = {Langmuir},
    author = {Pilat, D. W. and Papadopoulos, P. and Schäffel, D. and Vollmer, D. and Berger, R. and Butt, H.-J.},
    year = {2012},
    pages = {16812--16820},
}

@article{gao_how_2018,
    title = {How drops start sliding over solid surfaces},
    volume = {14},
    doi = {10.1038/nphys4305},
    number = {2},
    journal = {Nat. Phys.},
    author = {Gao, Nan and Geyer, Florian and Pilat, Dominik W. and Wooh, Sanghyuk and Vollmer, Doris and Butt, Hans-Jürgen and Berger, Rüdiger},
    year = {2018},
    pages = {191--196},
}

@article{ni_capillary_2018,
    title = {Capillary assembly as a tool for the heterogeneous integration of micro- and nanoscale objects},
    volume = {14},
    doi = {10.1039/C7SM02496G},
    number = {16},
    journal = {Soft Matter},
    author = {Ni, Songbo and Isa, Lucio and Wolf, Heiko},
    year = {2018},
    pages = {2978--2995},
}

@article{liu_capillary_2018,
    title = {Capillary {Assembly} of {Colloids}: {Interactions} on {Planar} and {Curved} {Interfaces}},
    volume = {9},
    shorttitle = {Capillary {Assembly} of {Colloids}},
    doi = {10.1146/annurev-conmatphys-031016-025514},
    journal = {Annu. Rev. Condens. Matter Phys.},
    author = {Liu, Iris B. and Sharifi-Mood, Nima and Stebe, Kathleen J.},
    year = {2018},
    pages = {283--305},
}

@book{israelachvili_intermolecular_2011,
    address = {Boston},
    title = {Intermolecular and {Surface} {Forces}},
    isbn = {978-0-12-391927-4},
    doi = {10.1016/C2011-0-05119-0},
    publisher = {Academic Press},
    author = {Israelachvili, Jacob N.},
    year = {2011},
}

@article{wong_bioinspired_2011,
    title = {Bioinspired self-repairing slippery surfaces with pressure-stable omniphobicity},
    volume = {477},
    doi = {10.1038/nature10447},
    number = {7365},
    journal = {Nature},
    author = {Wong, Tak-Sing and Kang, Sung Hoon and Tang, Sindy K. Y. and Smythe, Elizabeth J. and Hatton, Benjamin D. and Grinthal, Alison and Aizenberg, Joanna},
    year = {2011},
    pages = {443--447},
}

@article{david_smith_droplet_2013,
    title = {Droplet mobility on lubricant-impregnated surfaces},
    volume = {9},
    doi = {10.1039/C2SM27032C},
    number = {6},
    journal = {Soft Matter},
    author = {David Smith, J. and Dhiman, Rajeev and Anand, Sushant and Reza-Garduno, Ernesto and Cohen, Robert E. and McKinley, Gareth H. and Varanasi, Kripa K.},
    year = {2013},
    pages = {1772--1780},
}

@article{naga_direct_2024,
    title = {Direct visualization of viscous dissipation and wetting ridge geometry on lubricant-infused surfaces},
    volume = {7},
    doi = {10.1038/s42005-024-01795-3},
    number = {1},
    journal = {Commun. Phys.},
    author = {Naga, Abhinav and Rennick, Michael and Hauer, Lukas and Wong, William S. Y. and Sharifi-Aghili, Azadeh and Vollmer, Doris and Kusumaatmaja, Halim},
    year = {2024},
    pages = {1--12},
}

@article{naga_drop_2025,
    title = {Drop {Friction} and {Failure} on {Superhydrophobic} and {Slippery} {Surfaces}},
    volume = {19},
    doi = {10.1021/acsnano.5c01142},
    number = {20},
    journal = {ACS Nano},
    author = {Naga, Abhinav and Scarratt, Liam R. J. and Neto, Chiara and Papadopoulos, Periklis and Vollmer, Doris},
    year = {2025},
    pages = {18902--18928},
}

@article{hauer_wetting_2024,
    title = {Wetting on silicone surfaces},
    volume = {20},
    doi = {10.1039/D4SM00346B},
    number = {27},
    journal = {Soft Matter},
    author = {Hauer, Lukas and Naga, Abhinav and M. Badr, Rodrique G. and T. Pham, Jonathan and Y. Wong, William S. and Vollmer, Doris},
    year = {2024},
    pages = {5273--5295},
}

@article{quere_wetting_2008,
    title = {Wetting and {Roughness}},
    volume = {38},
    doi = {10.1146/annurev.matsci.38.060407.132434},
    number = {1},
    journal = {Annu. Rev. Mater. Res.},
    author = {Quéré, David},
    year = {2008},
    pages = {71--99},
}

@article{gresham_comparing_2025,
    title = {Comparing methods for preparing slippery liquid-like polydimethylsiloxane coatings},
    doi = {10.1038/s41596-025-01253-6},
    journal = {Nat. Protoc.},
    author = {Gresham, Isaac J. and Barrio-Zhang, Hernán and Cho, Jae Hyung and Khatir, Behrooz and Wells, Gary G. and Golovin, Kevin and McHale, Glen and Neto, Chiara},
    month = oct,
    year = {2025},
    pages = {1--21},
}

@article{daniel_origins_2018,
    title = {Origins of {Extreme} {Liquid} {Repellency} on {Structured}, {Flat}, and {Lubricated} {Hydrophobic} {Surfaces}},
    volume = {120},
    doi = {10.1103/PhysRevLett.120.244503},
    number = {24},
    journal = {Phys. Rev. Lett.},
    author = {Daniel, Dan and Timonen, Jaakko V. I. and Li, Ruoping and Velling, Seneca J. and Kreder, Michael J. and Tetreault, Adam and Aizenberg, Joanna},
    year = {2018},
    pages = {244503},
}

@article{fadeev_trialkylsilane_1999,
    title = {Trialkylsilane {Monolayers} {Covalently} {Attached} to {Silicon} {Surfaces}:  {Wettability} {Studies} {Indicating} that {Molecular} {Topography} {Contributes} to {Contact} {Angle} {Hysteresis}},
    volume = {15},
    shorttitle = {Trialkylsilane {Monolayers} {Covalently} {Attached} to {Silicon} {Surfaces}},
    doi = {10.1021/la981486o},
    number = {11},
    journal = {Langmuir},
    author = {Fadeev, Alexander Y. and McCarthy, Thomas J.},
    year = {1999},
    pages = {3759--3766},
}

@article{chen_omniphobic_2023,
    title = {Omniphobic liquid-like surfaces},
    volume = {7},
    doi = {10.1038/s41570-022-00455-w},
    number = {2},
    journal = {Nat. Rev. Chem.},
    author = {Chen, Liwei and Huang, Shilin and Ras, Robin H. A. and Tian, Xuelin},
    year = {2023},
    pages = {123--137},
}

@article{krumpfer_contact_2010,
    title = {Contact angle hysteresis: a different view and a trivial recipe for low hysteresis hydrophobic surfaces},
    volume = {146},
    issn = {1359-6640},
    shorttitle = {Contact angle hysteresis},
    doi = {10.1039/b925045j},
    journal = {Faraday Discuss.},
    author = {Krumpfer, Joseph W. and McCarthy, Thomas J.},
    month = jul,
    year = {2010},
    pages = {103--111},
}

@article{deng_direct_2018,
    title = {Direct measurement of the contact angle of water droplet on quartz in a reservoir rock with atomic force microscopy},
    volume = {177},
    doi = {10.1016/j.ces.2017.12.002},
    journal = {Chem. Eng. Sci.},
    author = {Deng, Yajun and Xu, Lei and Lu, Hailong and Wang, Hao and Shi, Yongmin},
    year = {2018},
    pages = {445--454},
}

@article{al_harraq_microplastics_2021,
    title = {Microplastics through the {Lens} of {Colloid} {Science}},
    volume = {2},
    doi = {10.1021/acsenvironau.1c00016},
    number = {1},
    journal = {ACS Environ. Au},
    author = {Al Harraq, Ahmed and Bharti, Bhuvnesh},
    month = sep,
    year = {2021},
    pages = {3--10},
}

@article{prajapati_microplastic_2022,
    title = {Microplastic properties and their interaction with hydrophobic organic contaminants: a review},
    volume = {29},
    shorttitle = {Microplastic properties and their interaction with hydrophobic organic contaminants},
    doi = {10.1007/s11356-022-20723-y},
    number = {33},
    journal = {Environ. Sci. Pollut. Res.},
    author = {Prajapati, Archana and Narayan Vaidya, Atul and Kumar, Asirvatham Ramesh},
    year = {2022},
    pages = {49490--49512},
}

@article{ma_ultra-thin_2021,
    title = {Ultra-thin self-healing vitrimer coatings for durable hydrophobicity},
    volume = {12},
    doi = {10.1038/s41467-021-25508-4},
    number = {1},
    journal = {Nat. Commun.},
    author = {Ma, Jingcheng and Porath, Laura E. and Haque, Md Farhadul and Sett, Soumyadip and Rabbi, Kazi Fazle and Nam, SungWoo and Miljkovic, Nenad and Evans, Christopher M.},
    year = {2021},
    pages = {5210},
}

@article{karakaya_silica_2026,
    title = {Silica {Sol}–{Gel} {Coatings} for {Solar} {Panels}: {Drop} {Friction} and {Particle} {Adhesion}},
    volume = {18},
    shorttitle = {Silica {Sol}–{Gel} {Coatings} for {Solar} {Panels}},
    doi = {10.1021/acsami.5c22794},
    number = {11},
    journal = {ACS Appl. Mater. Interfaces},
    author = {Karakaya, Tarik and Albarqawi, Sa’id and Sabath, Franziska and Sharifi-Aghili, Azadeh and Yavuz, Emre and Vollmer, Doris},
    year = {2026},
    pages = {17073--17082},
}

@article{wilson_effectiveness_2017,
    title = {The effectiveness of dust mitigation and cleaning strategies at {The} {National} {Archives}, {UK}},
    volume = {24},
    doi = {10.1016/j.culher.2016.09.004},
    journal = {J. Cult. Herit.},
    author = {Wilson, Helen and VanSnick, Sarah},
    year = {2017},
    pages = {100--107},
}

@article{qin_mechanisms_2003,
    title = {Mechanisms of particle removal from silicon wafer surface in wet chemical cleaning process},
    volume = {261},
    issn = {0021-9797},
    doi = {10.1016/S0021-9797(03)00053-5},
    number = {2},
    journal = {J. Colloid Interf. Sci.},
    author = {Qin, Kuide and Li, Yongcheng},
    year = {2003},
    pages = {569--574},
}

@article{panat_electrostatic_2022,
    title = {Electrostatic dust removal using adsorbed moisture–assisted charge induction for sustainable operation of solar panels},
    volume = {8},
    doi = {10.1126/sciadv.abm0078},
    number = {10},
    journal = {Sci. Adv.},
    author = {Panat, Sreedath and Varanasi, Kripa K.},
    year = {2022},
    pages = {eabm0078},
}

@inproceedings{sayyah_mitigation_2013,
    title = {Mitigation of soiling losses in concentrating solar collectors},
    doi = {10.1109/PVSC.2013.6744194},
    booktitle = {2013 {IEEE} 39th {Photovoltaic} {Specialists} {Conference} ({PVSC})},
    author = {Sayyah, Arash and Horenstein, Mark N. and Mazumder, Malay K.},
    year = {2013},
    pages = {0480--0485},
}

@article{parrott_automated_2018,
    title = {Automated, robotic dry-cleaning of solar panels in {Thuwal}, {Saudi} {Arabia} using a silicone rubber brush},
    volume = {171},
    doi = {10.1016/j.solener.2018.06.104},
    journal = {Sol. Energy},
    author = {Parrott, Brian and Carrasco Zanini, Pablo and Shehri, Ali and Kotsovos, Konstantinos and Gereige, Issam},
    year = {2018},
    pages = {526--533},
}

@article{yu_polymer-based_2023,
    title = {Polymer-based semiconductor wafer cleaning: {The} roles of organic acid, processing solvent, and polymer hydrophobicity},
    volume = {470},
    issn = {1385-8947},
    shorttitle = {Polymer-based semiconductor wafer cleaning},
    doi = {10.1016/j.cej.2023.144102},
    urldate = {2026-07-27},
    journal = {Chem. Eng. J.},
    author = {Yu, Seong Hoon and Jeon, Hayoung and Ko, Hyunki and Cha, Ji Hoon and Jeon, Soyeon and Jae, Mingyu and Nam, Geon-Hee and Kim, Kyoungsuk and Gil, Yeongjin and Lee, Kuntack and Chung, Dae Sung},
    year = {2023},
    pages = {144102},
}

@incollection{plummer_semiconductor_2023,
    title = {Semiconductor {Manufacturing}: {Cleanrooms}, {Wafer} {Cleaning}, {Gettering} and {Chip} {Yield}},
    booktitle = {Integrated {Circuit} {Fabrication}: {Science} and {Technology}},
    publisher = {Cambridge University Press},
    author = {Plummer, James D. and Griffin, Peter B.},
    year = {2023},
    pages = {111--158},
}

\pagebreak 
\newpage
\section{Acknowledgements}
A.N. and H.K. acknowledge the Cirrus UK National Tier-2 HPC Service at EPCC funded by the University of Edinburgh and EPSRC (EP/P020267/1) and the ARCHER2 UK National Supercomputing Service under the HPC access project ``Multiphysics Lattice Boltzmann Modelling of Self-Cleaning and Anti-Icing Phenomena on Functional Surfaces".

\section{Funding}

 This project was funded by a UK Engineering and Physical Sciences Research Council (EPSRC) National Fellowship in Fluid Dynamics (A.N., EP/X028410/2), an EPSRC Early Career Fellowship (H.K., EP/V034154/2), and the Max Planck – University of Twente Center for Complex Fluid Dynamics (D. V.). 

\section{Author contributions}
Conceptualization: A.N., D.V., H.K.; Methodology: A.N., D.V., H.K.; Software: A.N.; Validation: A.N, F.S., D.V., H.K.; Formal analysis: A.N., F.S.; Investigation: A.N., F.S.; Resources: A.N, F.S., D.V., H.K.; Data curation: A.N., F.S., D.V., H.K.; Writing - original draft: A.N.; Writing - review \& editing: A.N, F.S., D.V., H.K.; Visualization: A.N., F.S.; Supervision: A.N., D.V., H.K.; Project administration: A.N.; Funding acquisition: A.N., H.K., D.V.

\section{Competing interests}
All authors declare that they have no competing interests.

\section{Data, Code, and Materials Availability}

All data needed to evaluate and reproduce the results in the paper are present in the paper and/or the Supplementary Materials. The simulation code is available openly at https://doi.org/10.5281/zenodo.21823886. This study did not generate new materials.

\end{document}